\documentclass[aip, apl, floatfix]{revtex4}

\usepackage{amsfonts}
\usepackage{graphicx}

\DeclareMathAlphabet{\mathitb}{OT1}{cmr}{bx}{sl}

\begin{document}

\author{Paolo Michetti}
\email{michetti@physik.uni-wuerzburg.de}
\author{Bj\"orn Trauzettel}
\affiliation{Institute of Theoretical Physics and Astrophysics, University of W\"urzburg, D-97074 W\"urzburg, Germany}

\title[]{Devices with electrically tunable topological insulating phases}

\begin{abstract}
Solid-state topological insulating phases, characterized by spin-momentum locked edge modes,
provide a powerful route for spin and charge manipulation in electronic devices.    
We propose to control charge and spin transport in the helical edge modes 
 by electrically switching the topological insulating phase in a HgTe/CdTe double quantum well device.  
%
%
We introduce the concept of a topological field-effect-transistor and analyze possible applications to a spin battery, 
which also realizes a set up for an all-electrical investigation of the spin-polarization dynamics in metallic islands.
\end{abstract}

\maketitle


  The original prediction of the quantum spin Hall phase~\cite{kane2005a,kane2005b} has generated a renewed interest 
  in topological phases in solid state systems.
  In the following years, the realization of the topological insulator (TI) phase has been theoretically described and experimentally observed in 
  two-dimensional (2D) HgTe/CdTe quantum wells (QWs)~\cite{bernevig2006,konig2007}.

  1D helical edge modes are present at each boundary between a 2D TI and a normal insulator (NI), including the vacuum.
  Due to the topological protection against non-magnetic disorder~\cite{xu2006,Moore2007} and the spin-filtered nature of such 1D channels, 
  TI edge modes are extremely promising for spintronics.
  A crucial point, however, is to devise a reliable method allowing to control transport in these helical channels. 
  Several authors have proposed quantum interference and the Aharanov-Bohm effect as a possible route to control charge 
  and spin transport in 2D TIs~\cite{chu2009,maciejko2010,michetti2011,liu2011,citro2011,dolcini2011,krueckl2011,romeo2012}.
  However, the most straightforward route would be to change the TI into a NI and completely turn off the helical channel.
  Two recent proposals show that an electrically tunable TI phase can indeed be obtained in an InAs/GaSb type-II QW~\cite{liu2008} 
  and in a HgTe/CdTe double quantum well (DQW)~\cite{michetti2012}.

 \begin{figure}[bh]  
 \includegraphics[width=7.8cm]{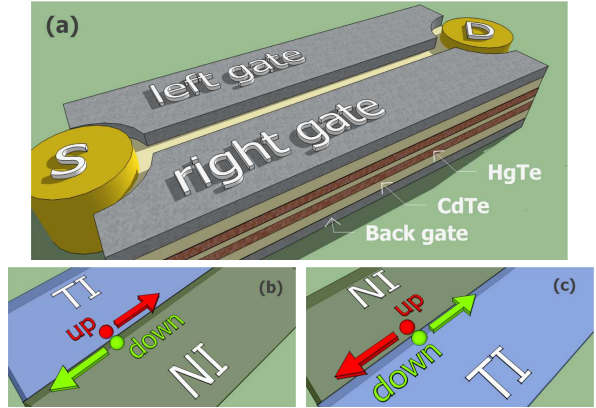}
 \caption{(Color online)
    (a) Isometric sketch of a HgTe/CdTe DQW device with a back gate and two distinct top gates (left and right).
    In the ON state, top gates induce a gate-bias domain leading to a TI/NI interface (channel) where helical edge modes are found. 
    Source (S) and drain (D) leads, placed along the interface between L and R top gates, collect charges from the edge modes.
    The lateral surface of the DQW is specifically treated to ensure negligible edge transport. 
    (b) Schematic description of a TI/NI interface for {\it direct} gate polarization with indication of the helical spin transport of edge states.
    (c) {\it Reverse} gate polarization leading to opposite spin transport of the channel.     
}
  \label{fig:device_art}
\end{figure}


HgTe DQWs are driven from NI to TI by the application of a inter-well potential bias $|V|>V_C$, 
where $V_C$ is a critical value on the order of the gap of the individual QWs~\cite{michetti2012}.    
As shown in Fig.~\ref{fig:device_art}(a), a left (L) and right (R) top gate are employed to generate an inter-well potential bias domain (PBD), 
with $V=V_L$ and $V=V_R$ in the L and R region, respectively.
The device is turned ON when the system realizes a TI/NI interface between the L and the R regions and 
helical edge modes run along the PBD line.
Hence, we distinguish between a {\it direct} gate polarization accodingly to $V_L>V_C>V_R$ and Fig.~\ref{fig:device_art}(b), where spin up electrons run from source to drain, 
and a {\it reverse} gate polarization $V_L<V_C<V_R$ with opposite spin transport properties [Fig.~\ref{fig:device_art}(c)].   
The device is turned OFF when the L and R regions belong to the same topological class, i.e. $V_L,V_R<V_C$ (both NI) or $V_L,V_R>V_C$ (both TI). 
Source and drain electrodes, which could for example be obtained by diffusing metallic atoms in the DQW system, 
lay close to the orifice between L and R top gates in order to collect electrons from the helical channel.

The design in Fig.~\ref{fig:device_art}(a) is studied to deal with a single helical channel (along the PBD), 
but helical edge states would also appear, if the DQW is in a TI phase, at the interface with the vacuum.
We will assume the lateral surfaces of the DQW to be specifically treated in order to impede charge transport in such edge modes. 
Indeed numerical simulations on quantum spin Hall systems~\cite{narayah2012} suggest that transport of an helical edge mode at a physical edge can be suppressed by locally doping 
with magnetic impurities of random or in-plane magnetization.
We will see that however such lateral edge modes, even when gapped out, play a relevant role in governing the spin properties of the system.

The low-energy physics of a single HgTe/CdTe QW is known to be captured by the Bernevig-Hughes-Zhang (BHZ) model~\cite{bernevig2006}, which
is, in first approximation, block-diagonal in the Kramers partner (spin) degree of freedom.
For the spin up block the BHZ Hamiltonian reads~\cite{bernevig2006}
\begin{eqnarray}
   h(\mathitb k)=  \left(C-D~\mathitb k^2\right)\mathbb I + \left(\begin{array}{cc} M-B~\mathitb k^2 & A \left(k_x+ik_y\right)\\
                                                                     A \left(k_x-ik_y\right) & -M+B~ \mathitb k^2                                                  
                                                   \end{array}\right),
\label{eq:H0}
\end{eqnarray}
written in the basis of the low energy subbands $\{$~$|E1,+\rangle$,~$|H1,+\rangle$~$\}$.
Conduction and valence bands of the TI for the spin up block are readily obtained by Eq.~(\ref{eq:H0}).
Spin down eigenstates can then be obtained by applying the time reversal operator $\hat{T}=-i \hat{K}$, with $\hat{K}$ the complex conjugation operator.
We adopt $A=375$~meV~nm, $B=-1.120$~eV~${\rm nm}^2$ 
and $D=-730$~meV~${\rm nm}^2$, which have been estimated by a comparison with the $8 \times 8$ Kane Hamiltonian~\cite{novik2005}, 
and assume $C=0$ without loss of generality. 
The Dirac rest mass $M$ depends on the QW thickness $d$.
For $d>6.3$~nm ($M<0$) the system realizes a TI and the real space solution of Eq.~(\ref{eq:H0}) allows for helical edge states at any TI/NI boundary~\cite{bernevig2006}.

The BHZ model has also been generalized to the case of a HgTe DQW~\cite{michetti2012}, 
where the two QWs are separated by a thin spacing layer of thickness $t$ allowing for tunneling.
The Hamiltonian for the spin up block has then the following structure 
\begin{equation}
    H_{DQW,\uparrow}(\mathitb k)\hspace{-0.05cm} =\hspace{-0.05cm} \left(\begin{array}{cc}
         h(\mathitb k)+\frac{V}{2} & h_T(\mathitb k) \\
         h_T(\mathitb k) & h(\mathitb k) -\frac{V}{2}  
      \end{array}\right), 
\label{eq:H_DQW}
\end{equation} 
where the tunneling Hamiltonian $h_T(\mathitb k)$ is given by
\begin{equation}
    h_T(\mathitb k) =  \left(\begin{array}{cc}
         \Delta_{E1} & \alpha \left(k_x+ik_y\right) \\
         \alpha \left(k_x-ik_y\right) & \Delta_{H1}  
      \end{array}\right), 
\label{eq:h_T}
\end{equation} 
omitting terms of order $O(k^2)$.
Numerical estimation of the tunneling parameters~\cite{michetti2012} indicates that $\Delta_{E1}\gg \Delta_{H1}$ so that  $\Delta_{H1}$ can be neglected.
Moreover $\Delta_{E1}$ and $\alpha$ can be taken constant as a function of $\mathitb k$.
Indicative estimates are reported in Table~\ref{table1} for $t$ ranging from $4$ to $7$~nm. 
%
\begin{table}[h]
\begin{tabular}{c|ccccccccc|}
   $t$~ (nm) &~ & ~~4~~ & ~~5~~ & ~~6~~ & ~~7~~ \\
  \hline
  ~$\Delta_{E1}$ (meV) &~ & 22 & 12 & 6 & 3.4  \\
  ~~~$\alpha$ (meV nm)&~ &  18 & 10 & 5 & 3 
 \end{tabular}
 \caption{Tunneling parameters $\alpha$ and $\Delta_{E1}$ for a HgTe/CdTe DQW, calculated at $k_x=k_y=0$ for several values of the spacing layer  thickness $t$~\cite{michetti2012}.}
 \label{table1}
\end{table}

\begin{figure}[bth]  
  \includegraphics[width=7.5cm]{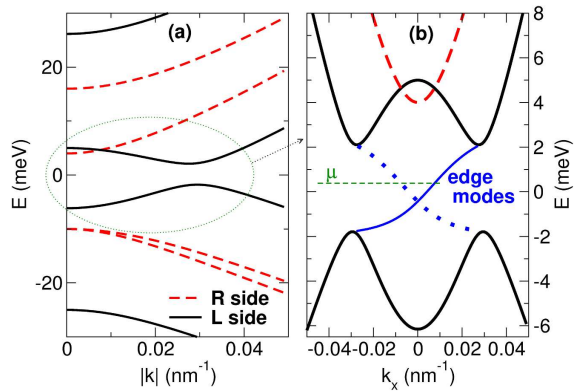}
  \caption{(Color online)
    (a) 2D bulk dispersion curves for the left (L) and right (R) regions of the DQW device, which are in the TI and in the NI state, respectively.  
    Bulk dispersion curves are degenerate in the spin degree of freedom. 
    (b) Detail of the low energy spectrum with a plot of the 1D helical edge modes at the TI/NI boundary (the PBD line). 
    Edge states in full thin lines and in dashed thin lines belong to opposite Kramers blocks (spin).
    We employ the following parameters: $t=5$~nm, $M=10$~meV, $V_L=30$~meV and $V_R=0$.
}
  \label{fig:disp}
\end{figure}
%

For definiteness, we consider a structure of two equal QWs with $M=10$~meV  (QW thickness $d\approx5.5$~nm), with a
thin inter-well spacing layer of thickness $t=5$~nm, ensuring a significant amount of tunneling, 
and consider a PBD where $V_L=30$~meV and $V_R=0$ in the {\it direct} gate polarization [Fig.~\ref{fig:device_art}(b)]. 
Fig.~\ref{fig:disp}(a) shows the 2D bulk band structure of the system, where full lines refer to the L side and dashed lines to the R side.
Note that the TI bandgap (L side) lays entirely in the NI bandgap (R side).
The bandgap in the TI phase develops due to the inter-layer tunneling interaction and indeed, keeping fixed $V=30$~meV, 
the TI gap is about $E_G\approx4$, $2$ and $1$~meV for a spacing layer $t=5$, $6$, $7$~nm, respectively.   
We solve now Eq.~(\ref{eq:H_DQW}) in real space, imposing the continuity of the wave function and its normal derivative through the PBD line.
When the device is ON, the system admits helical edge modes located at the TI/NI interface. 
The 1D dispersion curve of such helical edge states is shown in Fig.~\ref{fig:disp}(b), together with the merging bulk bands.

Let us now analyze ballistic transport in the helical edge states.
Due to their spin-filtering nature, it is useful to define separate chemical potentials $\mu_{\alpha,\sigma}$ for the two spin 
species ($\sigma= \uparrow$, $\downarrow$), with $\alpha=S$, $D$ for the source and drain contacts, respectively.
Chemical potentials are kept in the DQW gap, so that at sufficiently low temperature no charge is present in the bulk of the L and R regions.
For simplicity, we consider charge accumulation in the edge states (which are considerably extended) to have negligible effects.
The unidirectional charge current for the spin species $\sigma$ in the 1D helical channel is given by
\begin{eqnarray}
 I_\sigma^{(C)} =-\kappa_\sigma~\frac{e}{h} \int_{0}^{E_G} dE~ f(E-\mu_{X_\sigma,\sigma}) \approx -\kappa_\sigma~\frac{e}{h} \mu_{X_\sigma,\sigma}
\label{eq:I_sigma}
\end{eqnarray}
with $\kappa_\sigma=1$ ($-1$) and $X_\sigma=S$ ($D$) for $\sigma=\uparrow$ ($\downarrow$).
In the last passage we assumed $k_BT\ll E_G$,~$\mu_{X_\sigma,\sigma}$ and extended the integral to infinity. 
Note that $\mu$ is measured from the valence band edge of the TI region [Fig~\ref{fig:disp}(b)].
The helical currents satisfy $I_\uparrow^{(C)}\propto - \mu_{S,\uparrow}$ and $I_\downarrow^{(C)}\propto\mu_{D,\downarrow}$.

The device in Fig.~\ref{fig:device_art}(a) realizes the concept of a TI field effect transistor (FET), 
where charge transport in the 1D helical channel can be turned on and off by changing the topological phase of a part of the system. 
When the device is ON, the source-drain current in the 1D helical channel is 
\begin{equation}
 I_Q^{(C)} = I_\uparrow^{(C)} +I_\downarrow^{(C)}  = \frac{e}{h} \left[\mu_{D,\downarrow}-\mu_{S,\uparrow}\right],
\label{eq:I_Q}
\end{equation}
which gives $I_Q\approx0.1$~$\mu$A for $\mu_S=1$~meV and $\mu_D=3$~meV
(we assume spin-balanced leads with $\mu_{S,\sigma}=\mu_S$ and $\mu_{D,\sigma}=\mu_{D}$). 
This value does not significantly depend on temperature as long as $k_B T<E_{G}$,~$\mu_S$, $E_{G}-\mu_D$.

We estimate the charge currents in the 2D bulk bands as
\begin{equation}
I_Q^{(X, \pm)}\hspace{-0.1cm} = \hspace{-0.1cm} \frac{e~\alpha R^2}{\pi L_C\hbar} \hspace{-0.1cm}  \int\hspace{-0.1cm} dk\hspace{0.1cm}k\frac{{\rm d} E_{k}^{(X,\pm)}}{{\rm d} k} 
\left[ f(E\hspace{-0.1cm}-\hspace{-0.1cm}\mu_D)\hspace{-0.1cm} -\hspace{-0.1cm} f(E\hspace{-0.1cm}-\hspace{-0.1cm}\mu_S) \right],
\label{eq:I_Q_bulk}
\end{equation}
where $X=R$ or $L$, $\pm$ refers to conduction and valence bands, the factor $2$ accounts for the spin degeneracy, 
$L_C$ is the length of the channel and $R$ the typical size of the contacts.
$\alpha\approx \frac{R}{4\pi L_c}$ is the effective fraction of the electrons that has the right initial angle to propagate between the two contacts.

\begin{figure}[tb]  
  \includegraphics[width=7.5cm]{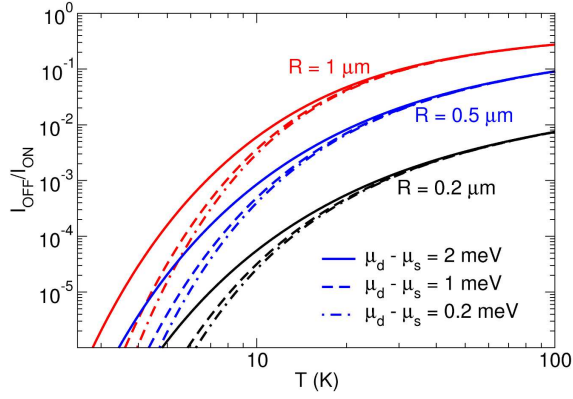}
  \caption{(Color online)
    $I_{\rm OFF}/I_{\rm ON}$ current ratio as a function of temperature of the topological FET in Fig.~\ref{fig:device_art}. 
    The channel length is $L_c=1$~$\mu$m, while we consider contacts of lateral size $R=0.2$, $0.5$, and $1$~$\mu$m.
}
  \label{fig:Iratio}
\end{figure}

We switch the device between the OFF state characterized by $V_L=V_R=0$ and 
the ON state with $V_L=0$ and $V_R=30$~meV, compatible with the situation described in Fig.~\ref{fig:disp}.
When the FET is OFF the helical channel is absent, $I_Q^{(C)}=0$, and therefore $I_{\rm OFF}= \sum_{X,\pm}I_Q^{(X, \pm)}$. 
The ON current is instead $I_{\rm ON}= I_Q^{(C)}+\sum_{X,\pm}I_Q^{(X, \pm)}$.
In Fig.~\ref{fig:Iratio}, we show the current ratio $I_{\rm OFF}/I_{\rm ON}$ as a function of temperature for different lateral size $R$ of the contacts.
%
%
It is clear that due to the narrow gap of the material small OFF/ON current ratios are only obtained at low temperature.
We note however that the bulk bands responsible for the OFF current have a 2D character, while both the contacts and the helical channel are quasi-1D.
Therefore the current ratio reduces for small $R$ (or for long channels $L_C$) 
where transport in the 1D edge states becomes more efficient than that in the 2D bulk bands.

A peculiarity of the topological FET described here is that a certain amount of disorder would be favorable for reducing the OFF/ON current ratio.
This stems from the topological protection against non-magnetic disorder of the helical channel, which would be little affected by the presence of inhomogeneities, 
while transport in the bulk bands (and therefore $I_{\rm OFF}$) would be suppressed .

The spin current in the helical channel of Fig.~\ref{fig:device_art} is given by
\begin{eqnarray}
  I_s^{(C)} = -\frac{\hbar}{2e} \left( I_\uparrow^{(C)}  - I_\downarrow^{(C)}  \right) = 
            \frac{1}{4\pi} \left[\mu_{S,\uparrow}+\mu_{D,\downarrow}\right].
 \label{eq:I_S}
 \end{eqnarray}
While the charge current Eq.~(\ref{eq:I_Q}) is evidently zero for the condition $\mu_{S,\uparrow}=\mu_{D,\downarrow}$, a net spin current would flow.
This is a persistent pure equilibrium spin current, which does not lead to spin accumulation~\cite{sonin2011}, 
because of the presence of helical edge states (even though they are gapped out) at the interface with the vacuum of the DQW.
Indeed a conduction gap in the lateral helical edge modes is identical to a perfect backscattering process plus a spin-flip.
Such backscattering processes lead to the following additional spin currents incoming on source and drain: 
\begin{equation}
  I_s^{(S)} =  \frac{1}{4\pi} 2 \mu_{S,\downarrow},\hspace{1cm}I_s^{(D)} = -\frac{1}{4\pi} 2 \mu_{D,\uparrow},
 \label{eq:I_s_add}
\end{equation}
and, as expected, in equilibrium ($\mu_{S,\sigma}=\mu_{D,\sigma}$) the total spin accumulation rate on the source (drain) 
due to helical edge modes $ I_s^{(S)}-I_s^{(C)}$ ($ I_s^{(D)}+I_s^{(C)}$) vanishes.

\begin{figure}[tb] 
    \includegraphics[width=5cm]{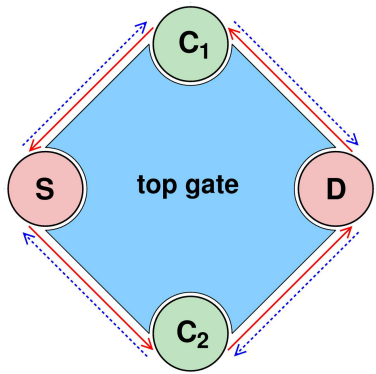}
  \caption{(Color online)
    Sketch of a spin battery realized with the exploitation of helical channels.
    Spin up (dashed line) and spin down (full line) spin filtered edge modes are shown at the boundary.  
}
  \label{fig:battery}
\end{figure}

As already noted in Ref.~\onlinecite{onoda2005, sonin2011}, spin accumulation exploiting the TI edge modes requires an applied source-drain bias.
Let us consider the setup shown in Fig.~\ref{fig:battery}, realized with an electrically tunable TI material.
We assume $C_1$ and $C_2$ to be mesoscopic metallic islands (MIs), while source and drain are macroscopic reservoirs with $\mu_{S,\uparrow}=\mu_{S,\downarrow}=\mu_S$ and 
$\mu_{D,\uparrow}=\mu_{D,\downarrow}=\mu_D$.
The system is described by three characteristic times: a spin-conserving relaxation time $\tau_E$, a current injection time $\tau_I$, and a spin relaxation time $\tau_s$~\cite{barnas2003}.
We express the particle number of the $\sigma$ spin species as $n_{S,\sigma}= N~\mu_{S,\sigma}$ 
(and similarly $n_{D,\sigma}= N~\mu_{D,\sigma}$), with $N=\frac{m R^2}{h \hbar}$ the constant 2D density of states and $R$ the typical dimension of the MIs.
The current injection time is then defined as $\tau_I = h N=\frac{m R^2}{\hbar}$, which for $m=0.1m_e$ and $R\approx1$~$\mu$m gives $\tau_I\approx1$~ns. 
For metallic islands of micrometer size generally $\tau_E\ll\tau_I$ and therefore we will assume intra-species relaxation to be instantaneous on the time scales of $\tau_I$. 
$\tau_s$ is in the range $10^{-9}-10^{-6}$~s~\cite{barnas2003}.
If $\tau_E\ll\tau_I\ll\tau_s$ the system is said to be in the {\it slow spin relaxation} limit.
We set up a minimal model, which is similar in spirit to what was used in Refs.~\onlinecite{barnas2000,barnas2003}, where the equilibration dynamics is described by master equations,
 and neglect for simplicity charging effects on the MIs. 
Hence, the charge and spin dynamics in the $C_2$ MI is governed by the following rate equation 
\begin{eqnarray}
    \tau_I~ \dot\mu_{C_2,\sigma} &=&   \mu_{X_\sigma} -  \mu_{C_2,\sigma} - \frac{\tau_I}{\tau_s}~\left[ \mu_{C_2,\sigma} -  \mu_{C_2,-\sigma} \right],
    \label{eq:dynamics2}
\end{eqnarray}
while the equation for $C_1$ is obtained by replacing $\mu_{C_2,\sigma}\leftrightarrow\mu_{C_1,-\sigma}$.
For spin-balanced leads with $\mu_S=-\mu_D=\delta\mu$ (for simplicity energy is now measured with respect to midgap) 
and the initially spin balanced MIs at $\mu_{C_1,\sigma}(t=0)=\mu_{C_2,\sigma}(t=0)=\mu_0$, 
Eq.~(\ref{eq:dynamics2}) has the solution  $\mu_{C_2,\sigma}(t)=\kappa_\sigma \tilde V + e^{-\frac{t}{\tau_I}}\left(\mu_0 - \kappa_\sigma \tilde V e^{-\frac{2t}{\tau_s}}  \right)$ and 
$\mu_{C_1,\sigma}(t) = \mu_{C_2,-\sigma}(t)$, with $\tilde V= \delta\mu~\frac{\tau_s}{2\tau_I+\tau_s}$.
In the slow spin relaxation limit, $\mu_{C_2,\sigma}\rightarrow \kappa_\sigma~\delta\mu$ exponentially with characteristic time $\tau_I$.
Opposite spin species accumulate on the two MIs with a spin-polarization factor $P= |\tilde V/E_0|$, assuming the conduction band edge of the MIs to be $E_0<-|\delta \mu|$.
Therefore $C_1$ and $C_2$ act like two opposite poles of a spin-battery.
When a normal conducting circuit closes on such poles a \emph{pure} spin current is supplied.
Note however that dissipation occurs in the spin battery [Fig.~\ref{fig:battery}] because of a finite source-drain voltage and a finite charge current.

The present configuration also realizes a set up where the dynamics of the spin densities on the MI can be studied by all-electrical means.
In fact, by switching OFF the helical channel, once that the current is in a steady state, we \emph{store} 
the spin densities on the two MIs, which then decays due to spin relaxation mechanisms.
In order to \emph{measure} the residual spin polarization $\mu_{C_1,\sigma}=-\mu_{C_2,\sigma}= \kappa_\sigma~\delta\mu(t)$ at the delay time $t$, 
we set $\mu_S=\mu_D=0$ and then turn ON again the helical channels.
The initial value of the source-drain charge current $I=-\frac{e}{h}~2\delta\mu(t)$ allows 
to extract the residual spin polarization in the MIs.
The time resolution is limited by the time it takes to switch the gate polarization (which does not involve any charging effects in the bulk). 

We neglected that in HgTe spin is not a properly conserved quantity
but rather the total angular momentum is, see for example the discussion in Ref.~\onlinecite{sonin2011}.
However, it can be shown that helical edge modes are spin polarized~\cite{michetti2011} (although spin is not perfectly aligned with $\hat z$) and Kramers partners are 
characterized by opposite spin. 
Moreover the polarization axis is fixed as long as Rashba terms are negligible~\cite{virtanen2012}.
We also assumed that edge modes always ideally sink into the electric contacts, however due to a non perfect overlap 
of the edge mode profile with source and drain, they could partially avoid the contacts 
complicating the physical picture.  
For a quantitative purpose all these aspects should be taken into account together with the charging effects on the helical channel and, 
when considered, on the mesoscopic metallic islands.

Many critical issues have still to be better understood before TIs can become a realistic solution for electronic devices.
One of the limiting factors of HgTe DQWs comes from the relatively large extent ($\lambda>100$~nm) of the edge states, 
which is due to the small band gap obtainable in that system.
It is therefore a crucial challenge to search for materials featuring larger TI gaps and well-localized edge states, like ultra-thin Bi films~\cite{murakami2006, wada2011} 
which has been proposed having $\lambda<10$~nm,
and devise a way to electrically tune the TI phase. 
Promising candidate materials include graphene doped with heavy adatoms~\cite{weeks2011}, silicene~\cite{liuCC2011, ezawa2012}, and 2D germanium~\cite{liuCC2011}.

\acknowledgements
We are grateful to G. Fiori and G. Iannaccone for helpful discussions and comments
We would like to thank the Deutsche Forschungsgemeinschaft, the European Science Foundation, 
as well as the Helmoltz Foundation for financial support.



\end{document}